# GEOMAGNETIC FIELD INTENSITY IN THE MIDDLE JURASSIC - OLIGOCENE

## A. Yu. Kurazhkovskii[1], N. A. Kurazhkovskaya[2], B.I. Klain[3]

[1, 2, 3]*Geophysical Observatory Borok, Schmidt Institute of Physics of the Earth of the Russian Academy of Sciences, Borok, Yaroslavl Region, 152742, Russian Federation*

[1]*E-mail:ksasha@borok.yar.ru*

[2]*E-mail:knady@borok.yar.ru*

[3]*E-mail:klain@borok.yar.ru*

**Abstract.** The present paper summarizes results of the studies on the intensity of geomagnetic field in the (167 - 23) Ma interval by sedimentary rocks of the Russian Plate and adjacent territories. The joint analysis of the data paleointensity obtained by sedimentary and thermomagnetized (from PINT12) rocks within this temporal interval is conducted. It is shown that the changes of the paleointensity were occurred chaotically. Alternating bursts and periods of quiet regime of the geomagnetic field are typical for intermittent processes and is a characteristic of the geological interval Jurassic – beginning of Paleogene. The distributions of the paleointensity corresponding to different intervals of geologic time were investigated. It is revealed that the cumulative distribution function (CDF) of the paleointensity values is best approximated by a power function. The indices of the power functions ($\alpha$) varied depending on geologic time intervals. The analysis of the paleomagnetic data suggests that the medium in which the geomagnetic field is generated is turbulent. Turbulence in the Earth's liquid core is enhanced in the Cretaceous compared with Jurassic and Paleogene.

**Keywords:** Jurassic- Cretaceous - Paleogene, paleointensity, generation of geomagnetic field, turbulence

## 1. Introduction

The study of the processes occurring in the central shell of our planet is one of the main directions in the field of geodynamics. The concepts on the evolutionary processes in the Earth's core are based on the indirect data, e. g. the changes in the characteristics of geomagnetic field may serve as indication of such processes. The investigation of statistical properties of the distribution of durations of polar intervals and of the values of intensity of ancient magnetic field allow for drawing conclusions on the state of medium in which the geomagnetic field was generated. The inversions of the geomagnetic field were the most significant events in its history. This is why the formation and use of the polarity scales for geodynamic researches were started earlier than the paleointensity data (*Cox*, 1968; *Khramov*, 1982).

Studies on the distributions of durations of the polar intervals were started simultaneously with publication of the first polarity scales. As it was shown in the papers (*Cox*, 1968; *Ruzmaikin and Trubikhin*, 1992) published in different times, the distributions of durations of polar intervals in the Cenozoic are well approximated by the exponential function. Statistical characteristics of the inversion regime at the beginning and at the end of the Cenozoic are significantly different (*Klain et al.*, 2009). This indicates that the processes in the Earth's core are non-steady and change during geological time.

The study on the paleointensity allows for analysis of the characteristic times and amplitude of its variations. Insufficiently detailed study of this characteristic of the geomagnetic field prevents widespread use of results of the paleointensity definitions for the study of processes in the Earth's core. The studies on the paleointensity of geomagnetic field in late Mesozoic - the Cenozoic were based mainly on the data obtained by analysis of thermomagnetized rocks. In several studies (*Valet*, 2003; *Biggin and Thomas*, 2003; *Heller et al.*, 2003; *Tarduno et al.*, 2006; *Tauxe and Yamazaki*, 2007) the distribution of the paleointensity values was analyzed. For example, *Heller et al.* (2003) revealed that depending on the intervals of the geological time the distribution of the paleointensity values could be either unimodal or bimodal. It has been suggested based on an analysis of distributions given in the same paper, that there are two sources of the geomagnetic field. At the same time, it was pointed out that available data are insufficient for confident conclusions about the work of a geodynamo.

Recently, new determinations of the paleointensity by sedimentary rocks of the Jurassic - Paleogene were published (*Kurazhkovskii et al.*, 2011, 2012, 2014; *Channell and Lanci*, 2014). However, the combined analysis of the paleointensity data of the Mesozoic - Cenozoic by sedimentary and thermomagnetized rocks were not performed yet.

The assessments of average and maximal values of the paleointensity of the same temporal intervals by sedimentary and thermomagnetized rocks may be slightly different (*Kurazhkovskii et al.*, 2007). As one of the reasons of such difference are the variations of the paleointensity and unevenness of sedimentation and volcanic activity. It was suggested (in some cases these suggestions were substantiated by geo-historical material) that changes in magmatic activity, ocean level and behavior of the geomagnetic field were interrelated. For example, according to *Seliverstov* (2001) the volcanism became more active at decline of the ocean level while the marine sedimentary layers on the continents were accumulated at high level of the ocean. At the decrease in the ocean level the frequency of the inversions of the geomagnetic field rose (*Milanovskii*, 1996). Activation of the effusive magmatism in the Mesozoic - Cenozoic was accompanied by the decrease in the frequency of inversions (*Larson and Olson*, 1991; *Dobretsov*, 1997) and increase in paleointensity (*Kurazhkovskii et al.*, 2007, 2010). This is why ancient sedimentary and magmatic rocks may carry different but mutually complementary information on the changes in the magnetic field. It is likely that validity of conclusions on the peculiarities of changes in the ancient geomagnetic field should arise at integral use of the data obtained by orientational and thermal remanent magnetization. The data on the behavior of paleointensity obtained by sedimentary rocks were not used yet at the discussions on the peculiar features of geomagnetic field in late Mesozoic - Cenozoic. This is why we attempted to partially fill this gap.

The goal of the present paper is to provide statistical analysis of the paleointensity data derived from studies on sedimentary and thermomagnetized rocks in the middle Jurassic – Paleogene.

## 2. The analyzed data

### 2.1. The paleointensity data by sediments

Results of the paleointensity determinations on sedimentary rocks were taken from (*Kurazhkovskii et al.*, 2011, 2012, 2014). The dynamics of the paleointensity of the middle Jurassic - Paleogene described in these studies is not continuous. For instance, these studies fully lack the paleointensity data of the Coniacian (*Kurazhkovskii et al.*, 2012) and Lutetian (*Kurazhkovskii et al.*, 2014. As the reason for the absence of the paleomagnetic information is the peculiarities of sediments accumulation. The summarized data obtained on the basis of the paleointensity determinations from 878 stratigraphic levels of the Jurassic – Paleogene deposits are shown in Fig. 1a. The average (for geological epochs) values of the paleointensity by sedimentary rocks are presented in Table 1.

**Table 1.** The average values of the paleointensity for geological epochs by sedimentary and thermomagnetized rocks.

| Age | Sediments | | | | Thermomagnetized rock | | | |
|---|---|---|---|---|---|---|---|---|
| | n | $H/H_0$ | $\sigma$ | $\alpha 95\%$ | n | $H/H_0$ | $\sigma$ | $\alpha 95\%$ |
| Oligocene | 46 | 0.56 | 0.35 | 0.11 | 112 | 0.58 | 0.34 | 0.07 |
| Eocene | 49 | 0.72 | 0.41 | 0.14 | 145 | 0.74 | 0.41 | 0.07 |
| Paleocene | 92 | 0.81 | 0.77 | 0.16 | 131 | 0.83 | 0.53 | 0.09 |
| Cretaceous 2 | 159 | 0.82 | 0.58 | 0.09 | 229 | 0.73 | 0.48 | 0.06 |
| Cretaceous 1 | 277 | 0.76 | 0.53 | 0.06 | 361 | 0.77 | 0.56 | 0.06 |
| Jurassic 3, 2 | 230 | 0.4 | 0.35 | 0.04 | 88 | 0.36 | 0.17 | 0.04 |

n is number of paleomagnetic determinations; $H/H_0$ is the average values of the paleointensity for the geological epochs, where $H_0 = 40$ µT; $\sigma$ is standard deviation; $\alpha 95\%$ is 95% confidence interval for the mean.

The intensity of geomagnetic field in the Jurassic - Paleogene varied in the complicated manner (chaotically) (Fig. 1a). However, the intervals differing in average values and amplitude of variation of the paleointensity can be detected in its behavior. The longest changes in the geomagnetic field had characteristic times of the order of geological epochs and periods. The average values of the paleointensity for geological epochs are significantly different (Table 1 and Fig. 1a). For example, the average value of the paleointensity in the Jurassic was $0.4H_0$ ($H_0$ is the value of modern geomagnetic field taken as 40 µT). During the middle and late Jurassic epochs the variations of paleointensity mainly had low amplitude. In early Cretaceous the amplitude of variations and mean values of paleointensity rose to $0.76H_0$. Inconsiderable growth of the paleointensity continued in late Cretaceous and in the early Paleogene. Within these time intervals the values of paleointensity averaged $0.82H_0$ and $0.81H_0$, respectively. At the end of Paleogene mean values and amplitude of variations of the paleointensity decreased comparing to the Cretaceous and beginning of the Paleogene. According to the changes in the amplitude of variations of the paleointensity (Fig. 1a) we conventionally divide the Paleogene into two parts: the beginning (Paleocene) and the end (Eocene – Oligocene).

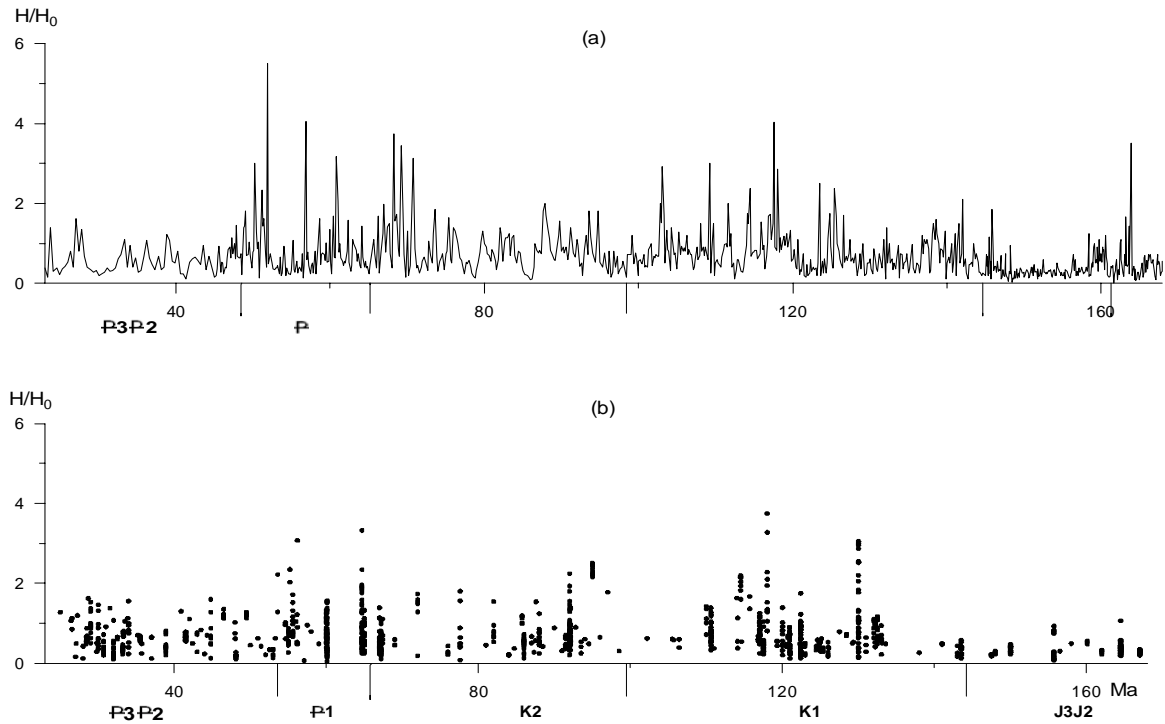

**Figure 1.** Dynamics of changes in geomagnetic field intensity within the (167 - 23) Ma interval (a) by sedimentary and (b) thermomagnetized rocks. The attribution of the fragments of paleointensity to the geological epochs according to the determination given in (*Bogachkin*, 2004; *Guzhikov et al.*, 2007; *Pimenov and Yampol'skaya*, 2008) are shown below abscissa.

In addition, the recurring geomagnetic events with characteristic times lesser than a geological age took place. For example, in the Cretaceous and the beginning of Paleogene considerable bursts of paleointensity (exceeding mean values by several times) took place. According to data in Fig. 1a and *Kurazhkovskii et al.* (2012), during one geological age one to two bursts of paleointensity may have occurred. Between these bursts the intervals of quiet geomagnetic field were noted. In the Jurassic and in the end of Paleogene the variations of the paleointensity mainly had relatively low amplitude (within $0.5H_0$). Thus, the data given here show that in the beginning of Cretaceous and in the middle of Paleogene considerable changes in the behavior of paleointensity took place.

*2.2 The paleointensity data by thermomagnetized rocks*

Results of the paleointensity determinations by thermomagnetized rocks were taken from the database PINT12 [http://earth.liv.ac.uk/pint/]. The description of this database is given by *Biggin et al.* (2010). The results of these determinations of the paleointensity in the interval (167 - 23) Ma are presented in Fig. 1b. In addition, the average values (for geological epochs) of the paleointensity obtained by thermomagnetized rocks are given in the Table 1. In the present study we used 1052 determinations of the paleointensity by thermomagnetized rocks. The paleointensity behavior obtained by sediments and thermomagnetized rocks exhibit some common regularities (Fig. 1). The amplitude of the paleointensity variations on sediments and thermomagnetized rocks increase from Jurassic to Cretaceous and decrease in the middle Paleogene. The data in Table 1 show that the average (for geological epochs) values and standard deviations of paleointensity obtained by sedimentary and thermomagnetized rocks vary by similar manner. At the same time, there is one obvious difference between these data massifs: the number of high values ($3H_0$ - $4 H_0$) of the paleointensity obtained by thermomagnetized rocks is lesser than those ones obtained by sedimentary rocks.

At the study of changes in the intensity of geomagnetic field we considered the behavior of the H, but not of the virtual dipole magnetic moment (VDM). This relates to several reasons. The statistics of the paleointensity distributions that will be studied below is in many respects determined by the presence of its high values. The high values of the paleointensity were obtained both by sedimentary and thermomagnetized rocks. The nature of high values of the paleointensity (bursts) unclear yet. Are the bursts phenomena of the regional or global scale? Consequently, we do not know the nature of the relationship between the paleointensity bursts and behavior of the Earth's magnetic dipole moment. The expediency and correctness of the translation of the H values in the dipole magnetic moment values does not seem obvious to us.

The PINT12 data show that the average values for geological epochs $H/H_0$, $V/V_0$ ($V_0$- modern geomagnetic dipole moment equal to $8*10^{22} Am^2$) and their dispersions coincide with good accuracy. This is why, in our opinion, it is not necessary to transform the data from H to VDM at the statistical studies of the evolutional changes in the geomagnetic field.

## 3. Statistical analysis of the paleointensity data

As we mentioned in the Introduction, the statistical characteristics of the paleointensity distributions were used repeatedly to study its changes on different intervals of the geologic time (see for example: *Biggin and Thomas*, 2003; *Heller et al.*, 2003). Distributions of the analyzed data are approximated by various functions, such as normal, lognormal, gamma function, etc. It should be noted that usually the choice of specific functions for the approximation of the empirical distributions lacks any theoretical substantiation. This is why there is a wide range of possibilities for the selection of functions to describe the obtained experimental distributions. An approach that uses "canonical" physical laws of the distribution is an alternative to the search of possible parametric distributions (*Sornette*, 2006; *Pisarenko and Rodk*in, 2007). The normal, exponential and power distributions attribute to "canonical". As emphasized by *Sornette* (2006); *Pisarenko and Rodkin* (2007) the use of this approach makes it possible to detect analogy with other phenomena. The physical interpretation of the results is also possible at using of this approach. In the case if the analyzed data are well described by one of the "canonical" distributions, we can consider this distribution exclusively. In the present study we analyze using sedimentary rocks data the behavior of "canonical" distributions of the paleointensity depending on intervals of the geological time.

### 3.1 Distribution of the paleointensity by sediments

The Fig. 2a shows the distribution of paleointensity values corresponding to various intervals of geological time (the middle - late Jurassic, Cretaceous and Paleogene). The modal values and amplitude of variations of the paleointensity (lengths of the "tails") were increasing from the Jurassic to Cretaceous (Fig. 2a). The paleointensity data are distributed asymmetrically relative to the modal values, i. e. these data do not obey the normal distribution. The approximations of accumulated distributions of the paleointensity (cumulative distribution function-CDF) by power and exponential functions are given in the right part of Fig. 2 (Fig. 2b). The $r_{power}$ and $r_{exp}$ are correlation coefficients between the paleointensity data and the power and exponential functions, respectively. In addition, indices of the power ($\alpha$) and the exponent ($\beta$) functions for each interval of geologic time are given. The approximation was performed within the intervals $x > x_0$, where $x_0$ stands for the modal value of paleointensity. As it is seen from Fig. 2b the values of paleointensity are best approximated by the power function. This is why further conclusions will be based on the results of approximation of the paleointensity data by the power function. The value of the power function index $\alpha$ varied depending on investigated intervals of the geological time (Fig. 2b). Index $\alpha$ has a maximum value at the approximation of the paleointensity CDF of the Cretaceous. It is likely, that the changes of the $\alpha$ index indicated the non- stationary processes associated with the generation of the geomagnetic field on the studied intervals of geologic time.

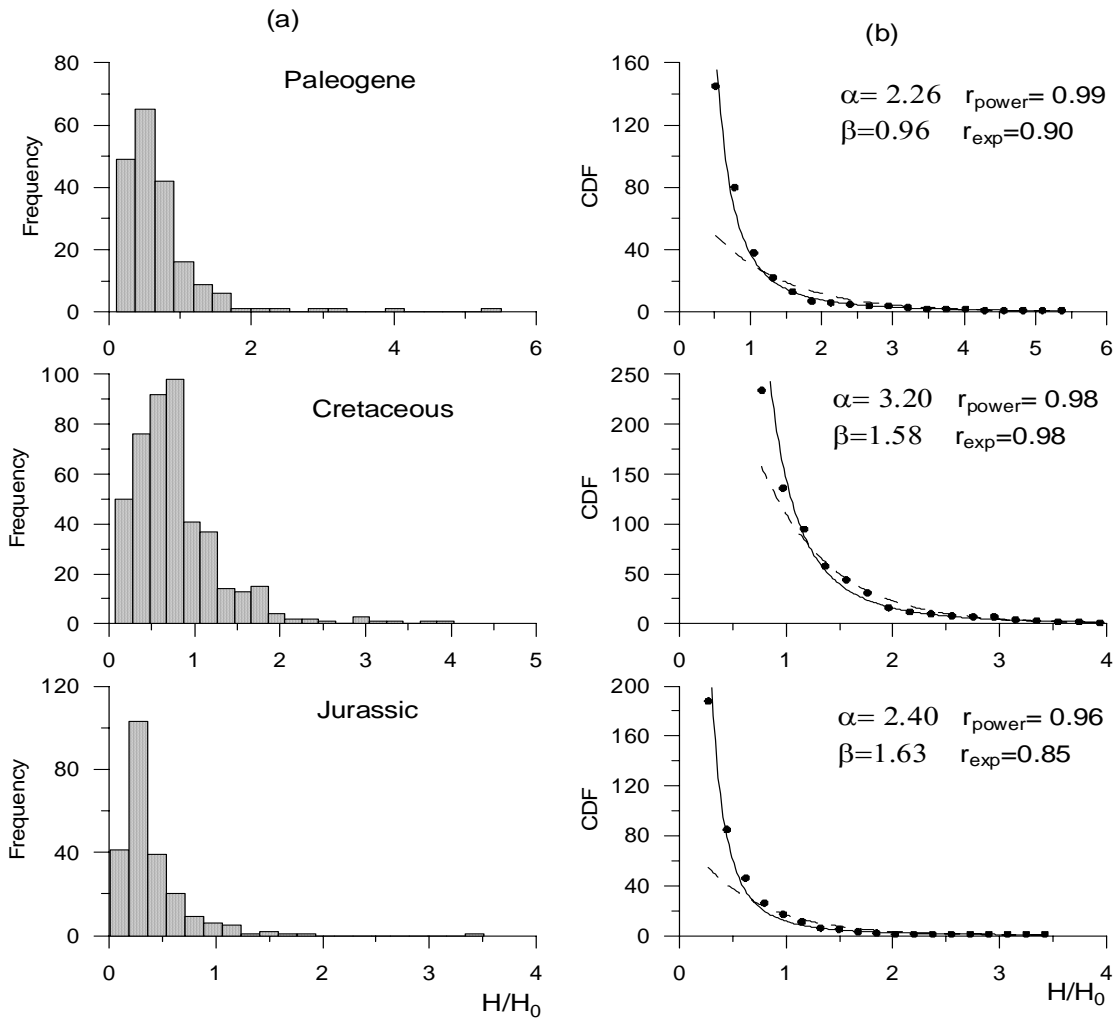

**Figure 2.** (a) Distributions of the paleointensity values by sediments for geological periods of the Jurassic - Paleogene and (b) their approximations by the power (solid line) and exponential (dashed line) functions. Circles denote the cumulative distribution function of the paleointensity (CDF). The $r_{power}$ and $r_{exp}$ are correlation coefficients between the experimental data and the approximating functions. The $\alpha$ and $\beta$ are indices of the power and exponential functions, respectively.

*3.2 Distribution of the paleointensity by thermomagnetized rocks*

The distribution of the paleointensity values in the middle Jurassic - Paleogene by thermomagnetized rocks from PINT12 is shown in Fig. 3a. By analogy with Fig. 2 the cumulative distributions of paleointensity (cumulative distribution functions - CDF) and their approximation by power and exponential functions (Fig. 3b) are given at the right part. As it follows from the histograms (Fig. 3a), the modal values and amplitude of variations of the paleointensity increased from the Jurassic to the Cretaceous and Paleogene. The most extensive "tails" in the distributions of paleointensity were observed in the Cretaceous and Paleogene. Thus, the distributions of the paleointensity values by sedimentary and thermomagnetized rocks lead to approximately the same conception of its changes (Fig. 2a and 3a).

The approximations of the paleointensity CDF by thermomagnetized (Fig. 3b) and sedimentary (Fig. 2b) rocks have some differences. For instance, the CDF on thermomagnetized rocks (of the Cretaceous) is somewhat better approximated by an exponential function than by a power function. At the same time, the dynamics of $\alpha$ indices on thermomagnetized and sedimentary rocks are about identical. The $\alpha$ minimum values are obtained at the approximation of distribution of the paleointensity in the Jurassic.

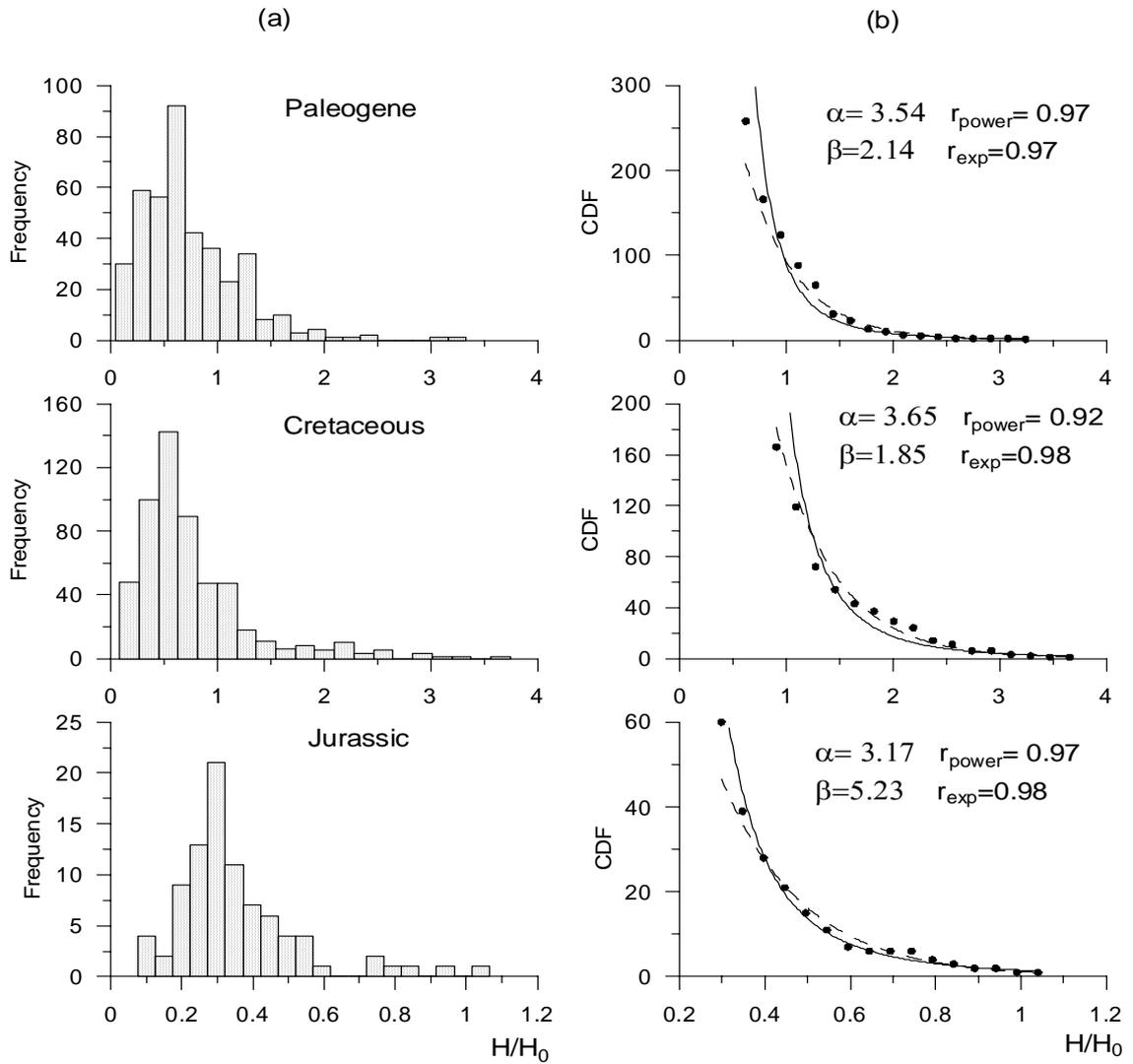

**Figure 3.** (a) Distributions of the paleointensity values of the Jurassic − Paleogene from PINT12, (b) their approximations by the power (solid line) and exponential (dashed line) functions. Circles denote the cumulative distribution function of the paleointensity (CDF). The $r_{power}$ and $r_{exp}$ are correlation coefficients between the experimental data and the approximating functions. The $\alpha$ and $\beta$ are indices of the power and exponential functions, respectively.

To obtain a more detailed view on the changes in paleointensity we have made approximation of the CDF according to the geological epochs. Approximation of the paleointensity distributions corresponding to geological epochs was performed by analogy with the data presented in Fig. 2b and 3b. The results of this approximation (value of the $\alpha$ index) are given in Table 2 and Fig. 4. The $\alpha$ dynamics obtained from sedimentary rocks do not change qualitatively depending on the resolution of the presented geomagnetic history (Fig. 4 and Table 2). For example, the $\alpha$ maximum value is obtained at the approximation of the CDF paleointensity of the early Cretaceous when a geomagnetic history is divided in accordance with the geological epochs. The $\alpha$ dynamics obtained by thermomagnetized rocks varies depending on the resolution of the partition of paleomagnetic history.

**Table 2.** The values of the indices of power (α) and exponential (β) functions approximating the paleointensity distributions (by sedimentary and thermomagnetized rocks) for different epochs.

| Epoch | Sediments | Thermomagnetized rock |
|---|---|---|
| End of the Paleogene | $\alpha = 2.50$ $r_{power} = 0.88$<br>$\beta = 3.20$ $r_{exp} = 0.99$ | $\alpha = 4.91$ $r_{power} = 0.87$<br>$\beta = 4.51$ $r_{exp} = 0.94$ |
| Beginning of the Paleogene | $\alpha = 1.85$ $r_{power} = 0.97$<br>$\beta = 0.87$ $r_{exp} = 0.92$ | $\alpha = 3.19$ $r_{power} = 0.96$<br>$\beta = 1.84$ $r_{exp} = 0.97$ |
| Late Cretaceous | $\alpha = 2.76$ $r_{power} = 0.96$<br>$\beta = 1.42$ $r_{exp} = 0.91$ | $\alpha = 2.36$ $r_{power} = 0.98$<br>$\beta = 1.76$ $r_{exp} = 0.98$ |
| Early Cretaceous | $\alpha = 3.18$ $r_{power} = 0.97$<br>$\beta = 1.56$ $r_{exp} = 0.95$ | $\alpha = 2.74$ $r_{power} = 0.93$<br>$\beta = 1.65$ $r_{exp} = 0.99$ |
| Late Jurassic | $\alpha = 2.89$ $r_{power} = 0.97$<br>$\beta = 3.19$ $r_{exp} = 0.95$ | $\alpha = 3.18$ $r_{power} = 0.97$<br>$\beta = 5.18$ $r_{exp} = 0.98$ |
| Middle Jurassic | $\alpha = 1.84$ $r_{power} = 0.92$<br>$\beta = 1.03$ $r_{exp} = 0.77$ | |

For instance, the maximum value of the α index corresponds to the Cretaceous when a geomagnetic history is divided according to the geological periods. At more detailed partition ("higher resolution") of a geomagnetic history the α minimum values correspond to epochs of the Cretaceous.

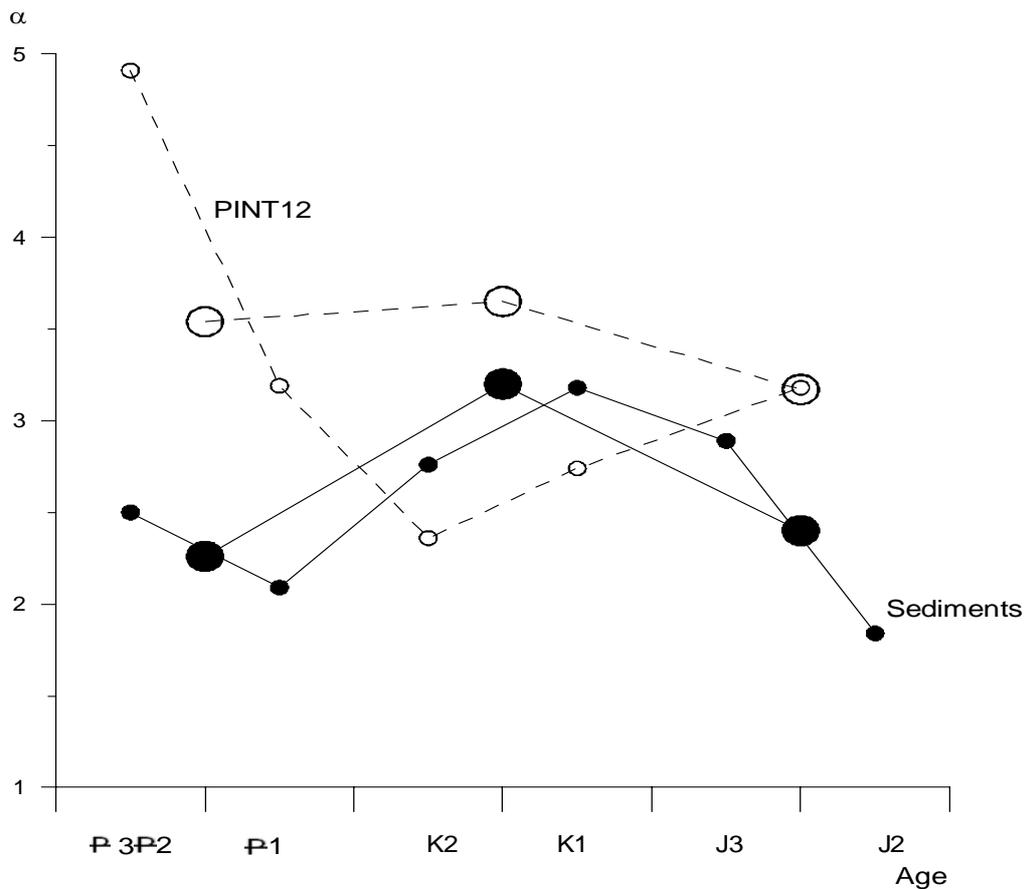

**Figure 4.** Dynamics of the power index *α* by sedimentary rocks (solid line) and thermomagnetized rocks (dashed line). The *α* index for geological periods and geological epochs by are denoted by large and small circles, respectively.

Thus, the behavior of α according to PINT12 data depends on the resolution of the geomagnetic history analysis. It is likely, this is due to insufficient amount of the data by thermomagnetized rocks.

## 4. Discussion

*4.1. Comparison of the results of the paleointensity determinations by analyses sedimentary and thermomagnetized rocks*

At the paleomagnetic studies two questions are always discussed: how correct and sufficient are the analyzed data to draw conclusions about the behavior of the geomagnetic field. The external convergence is the strongest argument in favor of the correctness of the paleomagnetic data, i.e. results of the paleomagnetic reconstructions should not depend on the method of obtaining the analyzed data and features of the genesis of geological objects on which they were received. The data in the Table 1 show that the mean values and dispersions of the paleointensity from sedimentary and thermomagnetized rocks coincide. This confirms the correctness of the determination of its average values.

As it was stated above, significantly more high values of the paleointensity were found by sedimentary than by thermomagnetized rocks. There could be two explanations for this difference. 1) The intensifications of magmatism associated with features of the paleointensity behavior. If we assume that during the bursts of H magmatic activity was weak, the high values of the paleointensity should rarely be detected by thermomagnetized rocks. Unfortunately it is impossible yet to verify the assumption on the relation of magmatism activations with the paleointensity behavior. The recent accuracy of dating does not allow for discussion of the problem of relations of magmatic activity with the paleointensity bursts. 2) The data obtained by analyses of thermomagnetized rocks present the geomagnetic history of the Cretaceous less completely than the data on sedimentary rocks. For example, PINT12 (Fig. 1b) contains 590 paleointensity determinations by thermomagnetized rocks of the Cretaceous period. The data in Fig. 1a represents 436 paleointensity determinations on the Cretaceous sediments. Number of the paleointensity determinations by analyses if sedimentary rocks are less abundant comparing to thermomagnetized rocks. However, according to isotopic dating only 88 paleointensity determinations from PINT12 differ in the time of its formation. The age of all 436 paleointensity determinations by analyses of sediments varies because the samples were collected from the different stratigraphic layers. Thus, the data obtained from sedimentary rocks give much more complete concept on the Cretaceous paleointensity than the data obtained from thermomagnetized rocks.

*4.2 Choice of the approximating function*

The analyzed data can be approximated either by power or exponential function (Fig. 2, 3 and Table 2). The choice of power-law distribution is insufficiently substantiated. In the context of the present paper the dynamics of the approximating function indices was more interesting than the dynamics pattern. At any choice of the approximating function its index changed depending on the geologic time intervals (Fig. 2, 3 and Table 2). Consequently, it is possible to conclude on qualitative changes in the behavior of the paleointensity based both on the dynamics of the power $\alpha$ index and on the dynamics of the exponent $\beta$ index.

At discussing of the changes of the paleointensity behavior it is necessary to pay attention to the general feature of the behavior of functions that approximate the paleointensity distributions obtained by analyses of sedimentary and thermomagnetized rocks. In the interval Jurassic – the beginning of the Paleogene the analyzed distributions are well approximated by a power function (Table 2). By the end of the Paleogene the paleointensity distributions by sedimentary and thermomagnetized rocks are better approximated by exponential function. Thus, both data massifs indicate that near the boundary of the Paleocene - Eocene significant changes in the behavior of the geomagnetic field took place.

*4.3 Changes of the turbulence of a medium in which the generation of the geomagnetic took place*

The results of the present study reveal the following regularities in the behavior of the ancient geomagnetic field: 1) changes in the paleointensity occurred chaotically; 2) the Jurassic – beginning of the Paleogene geological time interval is characterized by alternating bursts and periods of quiet regime of the geomagnetic field that is typical for intermittent processes (*Berzhe et al.*, 1991, *Platt et al.*, 1993); 3) the cumulative distribution functions of the paleointensity values in a given time interval in best approximated by a power function; 4) the indices of the power functions vary depending on the geological time, i.e. have different scaling dynamics.

In our opinion, these features of the behavior of the ancient geomagnetic field are a reflection of the turbulent processes occurring in the Earth's liquid core. Theoretically the role of turbulent processes in the generation of the geomagnetic field was discussed in the publications devoted to the work of the geomagnetic dynamo, e. g. in (*Braginsky and Roberts*, 1995; *Sokoloff*, 2004; *Reshetnyak*, 2008). The revealed regularities are based on the facts and indirectly confirm the effect of the turbulence on the generation of the geomagnetic field.

In addition, the regime of generation of the geomagnetic field in the interval Jurassic - beginning of the Paleogene is an intermittent process. The power distribution of the peaks intensity (bursts) is a distinctive feature of the intermittency. At that, the exponent reflects the state of the medium in which the bursts are formed (*Malinetskii and Potapov*, 2000). It is known that the intermittent processes are closely associated with the turbulence of the medium in which the burst regimes are generated (*Schuster*, 1988; *Berzhe et al.*, 1991, *Platt et al.*, 1993). The analyzed data consist of intervals with different scaling dynamics. This indicates the change of turbulence in the liquid core of the Earth over geologic time. In our opinion, the $\alpha$ index can serve as a quality characteristic of the level of turbulence (*Platt et al.*, 1993; *John et al.*, 2002). In accordance with the behavior of $\alpha$ we can conclude that the turbulence in the Earth's liquid core increased from the Jurassic to the Cretaceous and in the Paleogene decreased.

## 5. Conclusion

The analyzed materials revealed chaotic changes of paleointensity within the interval of (167 - 23) Ma. However several features of the paleointensity behavior specific for certain intervals of geologic time may be distinguished. The Jurassic – beginning of the Paleogene is characterized by alternating bursts and periods of quiet regime of the geomagnetic field that is typical for intermittent processes. Cumulative distribution functions of the paleointensity values in a given time interval is best approximated by a power law. No bursts of the paleointensity were revealed by the end of the Paleogene. By the end of the Paleogene CDF of the paleointensity is approximated better by an exponential function. The most significant changes in the behavior of the paleointensity occurred in close to the boundary of the Paleocene - Eocene. Indices of the approximating functions vary depending on the geological time intervals, i.e. have different scaling dynamics. Revealed regularities in the behavior of paleointensity indirectly confirm a significant effect of turbulence in the Earth's liquid core to generation of the geomagnetic field**.** In the Cretaceous the turbulence in Earth's liquid core was higher than in the Jurassic and the Paleogene**.**